\documentclass[%
 aip,
cp,  
 amsmath,amssymb,
 reprint,%
]{revtex4-2}

\usepackage{graphicx}
\usepackage{dcolumn}
\usepackage{bm}

\usepackage[utf8]{inputenc}
\usepackage[T1]{fontenc}
\usepackage{mathptmx} 

\begin{document}

\title{A Quasi-Local Inhomogeneous Dielectric Tensor For Arbitrary Distribution
Functions}

\author{S. J. Frank} 
 \email[Corresponding author: ]{frank@psfc.mit.edu}
\author{J. C. Wright}%
\author{P. T. Bonoli}
\affiliation{Massachusetts Institute of Technology Plasma Science and Fusion Center, 77 Massachusetts Ave, Cambridge, MA 02139 USA}

\date{\today} 

\begin{abstract}
Treatments of plasma waves usually assume homogeneity, but the parallel gradients ubiquitous in plasmas can modify wave propagation and absorption. We derive a quasilocal inhomogeneous correction to the plasma dielectric for arbitrary distributions by expanding the phase correlation integral and develop a novel integration technique that allows our correction to be applied in many situations and has greater accuracy than other inhomogeneous dielectric formulas found in the literature. We apply this dielectric tensor to the lower-hybrid current drive problem and demonstrate that inhomogeneous wave damping does not affect the lower-hybrid wave's linear damping condition, and in the non-Maxwellian problem damping and propagation should remain unchanged except in the case of waves with very large phase velocities. 
\end{abstract}

\maketitle
\section{Introduction \& Background}
Inhomogeneity is ubiquitous in plasmas. Most magnetic confinement fusion configurations, such as tokamaks, have a $\vec{B}\cdot\nabla \vec{B}$ inhomogeneity along their flux surfaces. However, despite tokamak plasmas being inhomogeneous, calculations of wave propagation and damping in tokamaks almost always are performed in the homogeneous slab limit. This approximation is generally considered to be "good", however, analyses of the effect of plasma inhomogeneity on wave damping have found that inhomogeneity \textit{can} meaningfully modify wave propagation and damping in tokamaks\cite{Faulconer1987,Smithe1988,Brambilla1994,Berry2016}. The effects of inhomogeneity are, in fact, of some importance to correctly capturing cyclotron resonance broadening and the ion Bernstien wave mode conversion in simulations of ion-cyclotron radio-frequency heating \cite{Brambilla1994,Brambilla1999}. Inhomogeneous effects have also been found to increase the Landau damping rate of waves with large phase velocities, and this has been used to suppress numerical pollution in simulations of shear Alfven waves \cite{Berry2016} and helicon waves\cite{Berry2016,Lau2018}. 

Calculations of the inhomogeneous plasma response have been generalized to arbitrary distribution functions in other work through calculations of the inhomogeneous perturbed current and quasilinear diffusion coefficient \cite{Lamalle1993,Lee2017,Catto2017,Catto2020,Catto2021a,Catto2021b}, but inhomogeneous dielectric tensors, which are much easier to quickly apply to calculations, have been limited to Maxwellian distribution functions\cite{Smithe1988,Berry2016}. Here, we generalize the work of Smithe \cite{Smithe1988} and Berry \cite{Berry2016} to general distribution functions and arbitrary resonances $n = 0 \rightarrow \pm \infty$. As this work is based on the quasilocal expansion technique, it does not guarantee global energy conservation in the same manner that techniques which consider the entirety of the particle motion in the tokamak do\cite{Lamalle1993,Catto2020,Catto2021a}. Despite this limitation, the quasilocal correction can still provide useful estimates of the importance of inhomogeneity in tokamaks and plasma experiments (especially in experiments with open field lines, such as the Large Plasma Device\cite{Gekelman2016}, where the quasilocal expansion should have good validity). We then will apply our inhomogeneous correction to lower-hybrid current drive (LHCD)\cite{Fisch1987} in Alcator C-Mod\cite{Greenwald2014} with both Maxwellian and non-Maxwellian electron distributions.
\section{Generalizing The Quasilocal Formulation}
We begin our derivation of the inhomogeneous correction to the dielectric tensor with the relativistic version of the hot-plasma dielectric tensor $\mathbf{\underline{\underline{\epsilon}}}$ defined in the Stix frame from \cite{StixPlasmaWaves}: 
\begin{equation} \label{eq:eps_start}
\epsilon_{ij} = \delta_{ij} + 2\pi\sum_s \sum_{n=-\infty}^{\infty} \frac{\omega_{ps}^2}{\omega^2}\int_0^\infty p_\perp dp_\perp \int_{-\infty}^\infty dp_\parallel
 \mathcal{Q}_{ij}^n(f_s,\vec{p},\vec{k},\omega)\int_{-\infty}^0 d\tau (-i\omega)e^{i(\omega-n\Omega_s-k_\parallel v_\parallel)\tau}, 
\end{equation} 
where $\mathcal{Q}_{ij}^n(f_s,\vec{p},\vec{k},\omega)$ is an expression dependent on the particular dielectric tensor element (the value of this function can be found in many texts on plasma waves \cite{StixPlasmaWaves,BrambillaPlasmaWaves} and at the moment is not important to us). In order to capture the effect of parallel inhomogeneity, we will consider the $d\tau$ integral in (\ref{eq:eps_start}), which when integrated in the homogeneous limit yields resonance condition:
\begin{equation}
\label{eq:resonance}
\mathcal{I}=\int_{-\infty}^{0}d\tau (-i\omega)e^{-i(\omega-n\Omega_s-k_\parallel v_\parallel)\tau} = \frac{\omega}{\omega-n\Omega_s - k_\parallel v_\parallel}.
\end{equation}
Here $\mathcal{I}$ tracks the relative interaction between particles and waves by measuring their phase correlation. We can include a correction to the phase correlation accounting for the effect of parallel inhomogeneity by Taylor expanding about $\tau = 0$ as follows:
\begin{equation}
\label{eq:taylor}
\mathcal{I} = \int_{-\infty}^{0}d\tau (-i\omega)e^{i(\alpha_n\tau+\beta_n\tau^2+...)} \simeq -i\omega e^{-\frac{i\alpha_n^2}{4\beta_n}}\int_{-\infty}^0 d\tau e^{i ( \frac{\alpha_n}{2\sqrt{\beta_n}}-\sqrt{\beta_n}\tau )^2},
\end{equation} 
where, dropping subscript $s$ for the time being, $\alpha_n = k_\parallel v_\parallel+n\Omega - \omega$ is the standard resonance condition and to leading order in a plasma with a $\vec{B}\cdot\nabla \vec{B}$ inhomogeneity \cite{Smithe1988,Berry2016,Itoh1985,GoodeThesis}:
\begin{equation}
\label{eq:betan}
\beta_n =
  \begin{cases}
    \frac{1}{2} v_\parallel^2 \frac{dk_\parallel}{dl} + \frac{1}{2}\frac{k_\parallel v_\perp^2}{B}\frac{dB}{dl} & \ n=0 \\ 
    -\frac{1}{2} v_\parallel \frac{n\Omega}{B}\frac{dB}{dl} & \ n>0,
  \end{cases}
\end{equation}
where $\beta_n$ is obtained from Taylor expanding accounting for the local variation in the $k_\parallel$, $v_\parallel$, and $\Omega$ along the field-line (\ref{eq:taylor}) with $d / dl$ indicating a derivative taken along the field line \cite{Smithe1988,Berry2016,Itoh1985}. The two terms in $\beta_0$ arise from variation in the wave phase and particle velocity respectively. The first term accounts for changes in phase velocity from varying $k_\parallel$. This occurs when the magnitude of $B, n, T$ or the B-field pitch angle change along the field line \cite{Berry2016}. The second term accounts for magnetic moment conservation that causes the parallel velocity of a particle to change as the magnitude of the magnetic field changes \cite{Smithe1988,Itoh1985}. For $\beta_{n>0}$ only the cyclotron broadening term is retained as it is much larger than other $\mathcal{O}(\tau^2)$ terms in the expansion \cite{Smithe1988}. Truncation at $\mathcal{O}(\tau^2)$, as we have done above, may not be entirely accurate for electrons or fast-ions as phase decorrelation by pitch angle scattering will be relatively slow \cite{Smithe1988,GoodeThesis}. In this case, it could be argued that higher order terms in the expansion resultant from effects such as orbit periodicity will become significant \cite{Catto2017,Catto2020,Catto2021a}. However, if the RF spectrum is sufficiently broad, higher order terms in the phase correlation expansion (\ref{eq:taylor}) will be quickly decorrelated by other components of the RF spectrum interacting with the particle. This broadband decorrelation argument is analogous to the phase-mixing argument made when deriving KE quasilinear diffusion \cite{KennelEngelmann1966}. Furthermore, if higher order terms are included, the phase correlation expansion approach becomes both analytically and computationally intractable. In this case, schemes such as basis function expansion \cite{Berry2016}, or kinetic approaches that more accurately account for the entirety of the particle's bounce motion \cite{Lamalle1993,Catto2017,Catto2020,Catto2021a} may be required. Returning to our derivation, integrating in $\varrho$ assuming $\textrm{Im}[\omega] > 0$ (this requirement can be argued from causality \cite{StixPlasmaWaves}) yields:
\begin{equation}
\label{eq:corrfun}
\mathcal{I}  = -2i\left( \frac{\omega}{\omega-n\Omega - k_{||}v_{||}} \right) \textrm{w}_n e^{-\textrm{w}_n^2} \int_{-\infty}^{i\textrm{w}_n} d\varrho e^{-\varrho^2}= -\left( \frac{\omega}{\omega-n\Omega - k_{||}v_{||}} \right)\textrm{w}_n\textrm{Z}(\textrm{w}_n),
\end{equation}
where  $\varrho = (-1)^{1/4}\left(\frac{\alpha_n}{2\sqrt{\beta_n}}-\sqrt{\beta_n}\tau \right)$, $\textrm{w}_n$ is a dimensionless parameter equal to $-\frac{\alpha_n (-1)^{1/4}}{2\sqrt{\beta_n}}$ (here we utilize the principle one-fourth root of negative one: $\frac{1}{\sqrt{2}} +\frac{i}{\sqrt{2}}$), and $\textrm{Z}(\textrm{w}_n)$ is the plasma dispersion function \cite{FriedConte}. In the limit of zero inhomogeneity, i.e. $\beta \rightarrow 0$, (\ref{eq:resonance}) is retrieved using the asymptotic expansion ($\textrm{w}_n \rightarrow \infty$) of $\textrm{w}_n\textrm{Z}(\textrm{w}_n)$. Equation (\ref{eq:corrfun}) can be substituted into the general hot plasma dielectric tensor (\ref{eq:eps_start}) in place of the standard result (\ref{eq:resonance}) to calculate the inhomogeneous correction to quasilocal wave propagation and damping.
\begin{figure}
\includegraphics[width=0.7\linewidth]{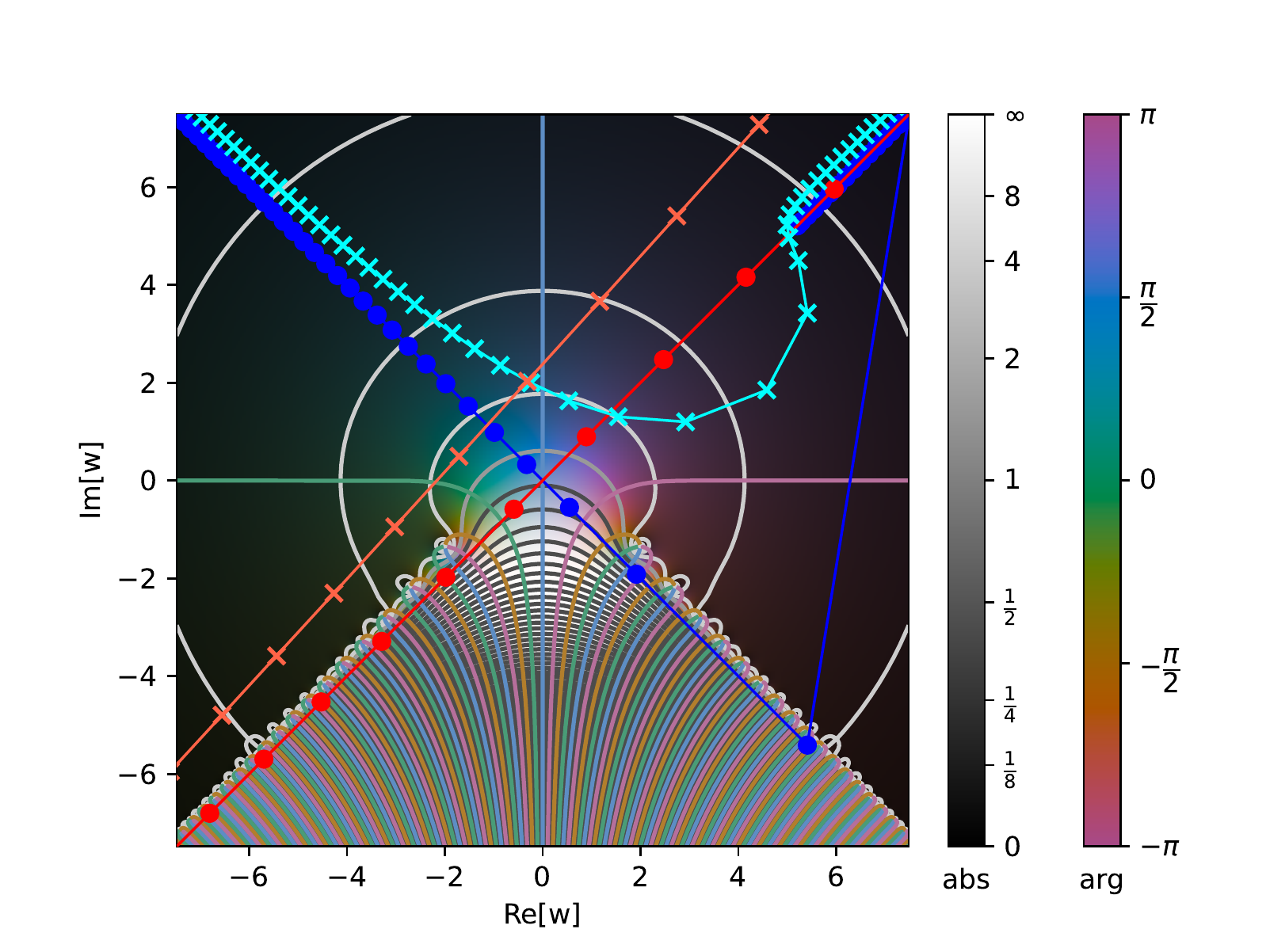}
\caption[Plots of the complex $\textrm{Z}$ function with integration paths overplotted using shifted and unshifted momentum contours.]{Example $\textrm{w}_0$ and $\textrm{w}_1$ integration paths, plotted in blue and red respectively, over contours of the $Z$-Function and its phase in the complex plane. The initial paths are plotted with dots and the shifted paths with crosses. Initially both $\textrm{w}_0$ and $\textrm{w}_1$ pass into the oscillatory lower midplane quadrant making numerical integration by most techniques impossible, but after being shifted to the modified $p_{\parallel}$ contour (\ref{eq:vcont}) neither enter the region removing oscillation in the integrand.}
\label{fig:zfunosc2}
\end{figure}

Problematically, when integrated over $p_\parallel$, $\textrm{Z}(\textrm{w}_n)$ in (\ref{eq:corrfun}) oscillates wildly as $\textrm{w}_n(v_\parallel)$ enters the lower midplane. This phenomena, demonstrated in Figure~\ref{fig:zfunosc2}, causes numerical integrals over $p_\parallel$ to converge extremely slowly or fail outright. Convergence difficulties from the introduction of inhomogeneity are, in fact, a general problem in the literature. For example, the modified Z-function integrals derived in Smithe and Berry \cite{Smithe1988,Berry2016} are not convergent under numerical integration. The root cause of the convergence problem is non-resonant wave-particle interactions. In the presence of inhomogeneity, particles at $\textit{all}$ velocities may have net interactions with incoming waves leading to oscillation of the integrand tracking the interactions. In previous work to obtain a solution the integral was truncated and only early time behavior is considered. Implementations of truncation, however, were generally ad-hoc and could fail when the plasma inhomogeneity became small. Here we present a novel contour integration approach that displaces the $p_\parallel$ integral into the complex plane to remove oscillations and also allows us to extend the inhomogeneous dielectric tensor to non-Maxwellian distributions. Using this contour, the dielectric for an arbitrary distribution function may be integrated to numerical precision without further approximation. Displacement of the $p_\parallel$ integral is desirable as it is bounded at $\pm \infty$. Because of this, the $p_\parallel$ integral can end anywhere in the complex plane when $\textrm{Re}[p_\parallel] = \pm\infty$ provided the distribution function being integrated has a valid analytic continuation and decays to 0 when $p_\parallel \rightarrow \pm\infty$. The contours typically used in the integral to obtain the dielectric \eqref{eq:eps_start} with inhomogeneous correction \eqref{eq:corrfun} are the following:
\begin{equation} \label{eq:vcont}
p_\parallel \rightarrow
  \begin{cases}
    p_\parallel-i\epsilon\,\textrm{Sign}\left[v_\parallel\frac{dk}{dl}\right] & \ n=0 \\ 
    p_\parallel-i\epsilon\,\textrm{Sign}\left[\frac{1}{B}\frac{dB}{dl}\right] & \ n>0
  \end{cases},
\end{equation}
where $\epsilon$ is a small fixed parameter. These contours divert the integration path away from the $\textrm{Z}$ function's oscillatory region in the lower midplane removing rapid oscillation from the integrand. It may at first appear that \eqref{eq:corrfun} introduces a pole in the integrand at the typical $\omega - k_\parallel v_\parallel - n \Omega \rightarrow 0$ resonance, or in situations when $\beta \rightarrow 0$ from $v \rightarrow 0$. However, the leading resonant denominator cancels with the $-\alpha_n$ in the numerator of the $\textrm{w}_n$ outside the $\textrm{Z}$ function, and it can be shown by asymptotic expansion of the $\textrm{Z}$ function for $v \rightarrow 0$ there are no poles in the integrand (in fact, this integral generally has no poles unless the chosen distribution function introduces one). This allows quick and accurate evalution of (\ref{eq:eps_start}) with inhomogeneous correction (\ref{eq:corrfun}) using simple Gaussian quadrature and the above contour \eqref{eq:vcont} (Note: for $n=0$ the $\pm i\epsilon$ zero crossing must also be integrated). An exaggerated example of these contours and how they divert the integration path can be seen in Figure~\ref{fig:zfunosc2}. This integration method allows us to robustly apply our inhomogeneous correction to any distribution function with an analytic continuation and completes the derivation here. 
\section{Application To LHCD}
Now, we will apply our inhomogeneous correction to a model problem to determine the effect of inhomogeneity on LHCD. Here, we take the non-relativistic limit and will not consider the inhomogeneous correction from velocity variation (this was done so we could validate our results to those found in Berry \cite{Berry2016}). The fully relativistic calculation including particle velocity variance will be saved for later work. Measurements of inhomogeneity in tokamaks using AORSA \cite{Berry2016} indicate that generally the peak $dk_\parallel/dl \lesssim 10$ m$^{-2}$. We will apply our correction to the hot-plasma dielectric to an Alcator C-Mod LHCD scenario examined in much previous modeling work ($\omega/2\pi = 4.6$ GHz, $n_e = 7\times10^{19}$ m$^{-3}$, $T_e = 2.3$ keV, $B_0 = 5.4$ T, and $dk_\parallel/dl = 10$ m$^{-2}$)\cite{SchmidtThesis,Wright2009,Wright2014,Frank2022a,Frank2022b}.
\begin{figure*}
\includegraphics[width=1\linewidth]{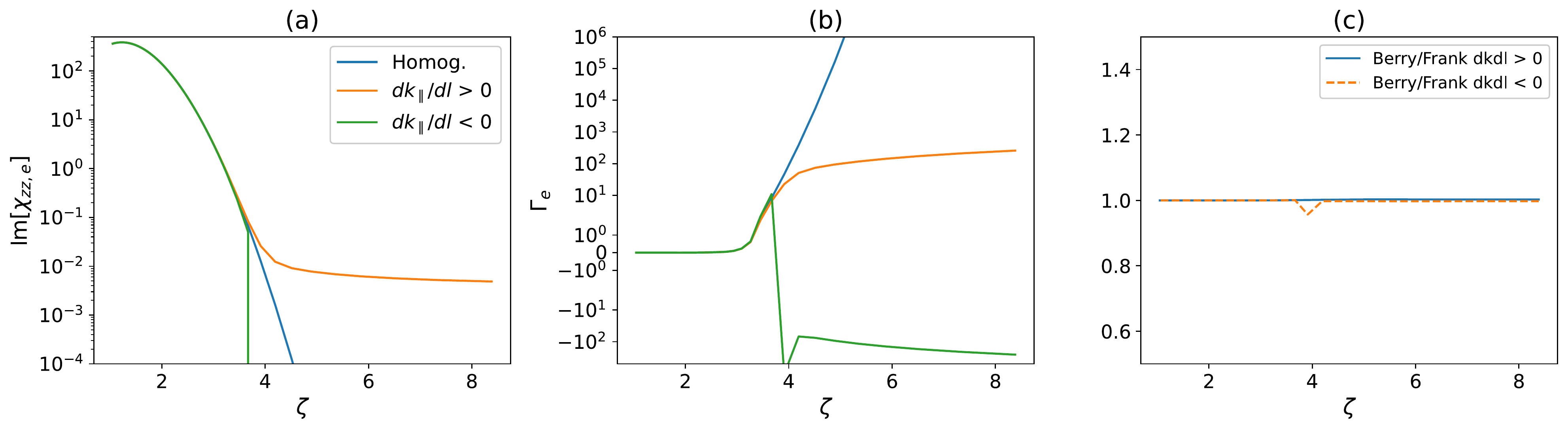}
\caption{(a) Plots of $\chi_{zz}$ versus $\zeta = \omega/k_\parallel v_{the}$, (b) the normalized damping parameter $\Gamma_e$ versus $\zeta$ comparing the homogeneous solution with the inhomogeneous solution for $dk_\parallel/dl > 0$ and $dk_\parallel/dl < 0$, and (c) ratio of the Berry dielectric \cite{Berry2016} and our dielectric for $dk_\parallel/dl > 0$ and $dk_\parallel/dl < 0$.}
\label{fig:chizzinhomog_max}
\end{figure*}

In the case of LHCD, we are interested in the modification of the $\chi_{zz,e}$ term which governs Landau damping and is the dominant term in determining propagation of the slow wave \cite{Wright2014b}. This can be written with our inhomogeneous correction as:
\begin{equation}\label{eq:inhomogchizz}
    \chi_{zz,e} = \mathcal{A} \int_0^\infty du \, u e^{-u^2} J_0^2\left(k_\perp \rho_e u\right) \int_{-\infty}^\infty dw \, \textrm{Sign}[w] we^{-w^2}\textrm{Z}[\textrm{w}_0]
\end{equation}
where $\mathcal{A}=2(-1)^{1/4}\sqrt{\frac{2}{\pi dk_{||}/dl}}\frac{\omega_{pe}^2}{\omega v_{the}}$ is a dimensionless factor, $u=v_\perp/v_{the}$, $\textrm{w}_0 = \frac{1}{2} v_\parallel^2 \frac{dk_\parallel}{dl}$, and $w=v_{||}/v_{the}$. We may integrate this along our modified velocity contours \eqref{eq:vcont} using Gaussian quadrature (replacing $p_\parallel$ with $w$ in the equation). Integrating \eqref{eq:inhomogchizz} for Alcator C-Mod like parameters ($n_e = 7\times10^{19}$ m$^{-3}$, $T_e = 2.3$ keV, $B_0 = 5.4$ T, $dk_\parallel/dl = 10$ m$^{-2}$ and $\omega/2\pi = 4.6$ GHz) we obtain the result shown in Figure~\ref{fig:chizzinhomog_max}. Our contour evaluation effectively resolved the oscillations in the integrand and allowed us to obtain the $\chi_{zz}$ for both positive and negative values of $dk_\parallel/dl$. We then used the LH dispersion relation \cite{Bonoli1984} to calculate normalized linear damping length $\Gamma_e = k_{\perp i} a$ and determine if the inhomogeneous correction could meaningfully change the linear Landau damping condition. This analysis, shown in Figure~\ref{fig:chizzinhomog_max}b, demonstrated that despite the profound difference in damping rate at large $\zeta=\omega/k_\parallel v_{the}$ from the introduction of inhomogeneity, wave damping remained unchanged in the region of linear damping $\zeta \sim 3$ and $\Gamma_e \le 1$ \cite{Bonoli1984}. At large $\zeta$, while the $\Gamma_e$ is much smaller when inhomogeneous damping is taken into account, $\Gamma_e$ is still far too large for inhomogeneity to have an observable impact on LH wave damping in tokamaks. The toroidal upshift mechanism will close the spectral gap long before the wave has propagated far enough to experience significant inhomogeneous damping \cite{Bonoli1981,Bonoli1982}. As a check, we reproduced our Maxwellian results with the Berry dielectric integral \cite{Berry2016} truncated at $\omega \tau = 110$ and obtained quantitative agreement with our dielectric integral. This result is shown in Figure~\ref{fig:chizzinhomog_max}.
\begin{figure}
\centering
\includegraphics[width=0.45\textwidth]{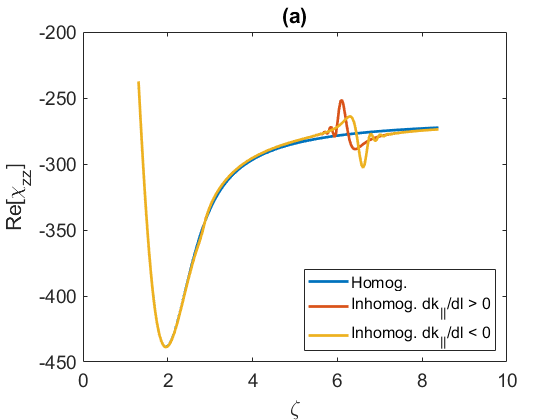}
\includegraphics[width=0.45\textwidth]{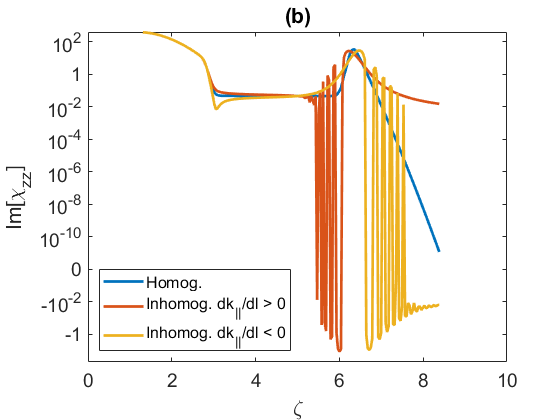}
\caption{Plots comparing the non-Maxwellian homogeneous and inhomogeneous $\chi_{zz}$ versus $\zeta$ with a Landau plateau distribution for $dk_\parallel/dl > 0$ and $dk_\parallel/dl < 0$. (a) The real part of  $\chi_{zz}$ with a Landau plateau distribution versus $\zeta$, and (b) the imaginary part of $\chi_{zz}$ for a Landau plateau distribution versus $\zeta$.}
\label{fig:inhomogNonMax}
\end{figure}

Now, let us consider the case of non-Maxwellian damping. To do this, we used the closed form solution to the non-Maxwellian Landau plateau distribution function from \cite{Karney1979,Frank2022b}. We applied the AAA algorithm \cite{Nakatsukasa2018} which creates an approximate numerical analytic continuation for arbitrary functions, to the non-Maxwellian distribution function. After this step, we integrated the dielectric with our approximate analytic continuation of the Landau plateau distribution along our complex velocity contour to find $\chi_{zz}$ as we did before in the Maxwellian case. The results of this calculation using Alcator C-Mod plasma parameters and a box $D(w)$ with $w_1 = 3$, $w_2 = 6$, and $D_0 = 1$ are shown in Figure~\ref{fig:inhomogNonMax}. When LH wave damping becomes non-Maxwellian, the differences between the inhomogeneous dielectric and the homogeneous dielectric are reduced relative to one another at low $\zeta$. However, inhomogeneity induces oscillation near the large $\zeta$ edge of the Landau plateau. This is a result of electrons at the Landau plateau edge, which previously could not interact with the LH waves as they are by definition outside of the wave spectrum (the edge of $D_{ql}$ has a one-to-one correspondence with the edge of the wave spectrum), interacting with the waves because the Landau resonance has become de-localized in velocity-space. This occurs in both the real and imaginary part of the dielectric, however, the oscillations in the real part are relatively small $\sim 15 \%$ and limited to $\zeta$ values near the plateau edge. Thus, most cases it should still be acceptable to use a Maxwellian homogeneous $\textrm{Re}[\chi_{zz}]$. There may be noticeable effects on wave damping for very high $\zeta$ waves, however, it is also possible these oscillations would disappear if wave-particle decorrelation mechanisms were added as they result from wave interactions with particles at large $\tau$.
\section{Conclusions}
We generalized previous work \cite{Faulconer1987,Smithe1988,Berry2016} to derive a novel form of the hot-plasma dielectric tensor for arbitrary relativistic distribution functions with an inhomogeneous correction. To evaluate the inhomogeneous dielectric, we developed a contour integration method. We then applied our inhomogeneous correction to evaluate the importance of inhomogeneity to LHCD in Alcator C-Mod in a simplified case. We found that inhomogeneous damping should have no effect on wave propagation or the linear Landau damping condition. Finally, we extended the inhomogeneous damping formulation to a non-Maxwellian distribution function with a Landau plateau. In this case, we found that the correction for inhomogeneous damping is generally small for all but the highest $\zeta$ waves. These waves may be able to interact with electrons at the edge of the plateau, that normally do not interact with waves, because of resonance broadening from the inhomogeneous correction. As parallel inhomogeneity becomes smaller with increasing aspect ratio and $R_0$, we anticipate that parallel inhomogeneity should not play a major role in tokamak LHCD experiments (our calculations were based on Alcator C-Mod which has a smaller $R_0$ than other experiments).  While the analysis here indicates that the effect of inhomogeneity on LH wave damping should not be large in most cases, it is important to realize that we have only applied a quasi-local inhomogeneous correction to the dielectric. To fully describe the effect of inhomogeneity on wave damping and propagation a more detailed analysis that considers the wave particle correlation over the entirety of the particle's bounce motion, like that performed by Catto or Lamalle \cite{Lamalle1993,Catto2017,Catto2020,Catto2021a,Catto2021b}, is required (this also resolves the non-positive definiteness of power transfer between waves and particles we see with our first-order correction). However, the results of these analyses are very difficult to apply to realistic problems.
\begin{acknowledgments}
The authors wish to thank P. J. Catto and I. H. Hutchinson for helpful discussions about this work.
\end{acknowledgments}
\nocite{*}
\bibliography{main}

\end{document}